\documentclass[a4paper,10pt]{article}\textheight 24.5cm \textwidth 17cm\voffset=-1.in\hoffset= - 1.in

\usepackage[dvips]{graphicx}
\usepackage{psfrag}
\usepackage{amsmath}
\usepackage{graphicx}
\usepackage{amsfonts}
\usepackage{amssymb}
\usepackage{mathtext}         %
\usepackage[T2A]{fontenc}     %
\usepackage[cp1251]{inputenc} %

\def\cP{{\mathcal{P}}}

\def\pd{\partial}

\def\tr{{\mathop{\rm tr}}}

\def\ve{\varepsilon}
\def\la{ \langle }
\def\ra{ \rangle }
\def\ti{ \tilde{t} }
\def\Fi{ \tilde{F} }
\def\ui{ \tilde{u} }
\def\Ri{ \tilde{R} }

\begin{document}


\vspace{0.5cm}

\begin{center}
\begin{LARGE}
{\Large \textbf{A remark on the three approaches to 2D Quantum gravity }}

\vspace{0.3cm}

\end{LARGE}

\vspace{0.5cm}

{\large A.~Belavin$^{a}$, M.~Bershtein$^{a,b}$, G.~Tarnopolsky$^a$ }

\vspace{.4cm} { \it
$^a$ Landau Institute for Theoretical Physics, 142432 Chernogolovka of Moscow Region, Russia\\

\vspace{0.3cm}
$^b$ Independent University of Moscow, 11 Bolshoy Vlasyevsky pereulok, 119002 Moscow, Russia\\
} \vspace{0.3cm} e-mail: belavin@itp.ac.ru, mbersht@gmail.com, hetzif@itp.ac.ru
\large

\vspace{1.0cm}
\begin{center}
   \textbf{Abstract  }
\end{center}

\begin{flushleft}
The one-matrix model is considered. The generating function of the correlation numbers is defined in such a way that this function coincides with the generating function of the Liouville gravity. Using the Kontsevich theorem we explain that this generating function is an analytic continuation of the generating function of the Topological gravity. We check the topological recursion relations for the correlation functions in the $p$-critical Matrix model.
\end{flushleft}

\vspace{.8cm}


\end{center}

\section{Introduction}
\large
There exist at least three approaches to the 2D
Quantum gravity namely the Liouville gravity (LG), the Matrix models (MM) and the
Topological gravity (TG). Details and references can be found in reviews e.g. \cite{rev1,rev2}

In this paper we consider the particular $p$-critical one-matrix model. The correlation functions are defined as derivatives of the free energy function $F(t_0,t_1,\ldots,t_{p-1})$ at some point. The key property of the free energy function of the  matrix model is the fulfilment of the string equation and the KdV equations. We consider the expansion of the free energy function at the particular point $t_0=\mu,\, t_1=t_2=\ldots=t_{p-1}=0$ and choose the particular solution of the string equation. We explain this choice of the boundary conditions in Section 2. This choice is determined by the agreement with the $(2,2p+1)$ Minimal Liouville gravity.

For the simplest case $p=2$ the coincidence between the correlation functions in both approaches is straightforward. In the general case the coincidence can be reached after a substitution of variables suggested in \cite{BZ}. Note, that the coincidence was checked for many cases \cite{BZ, BT,VBelavin} but wasn't proved rigorously, due to the fact that correlation functions in the Liouville theory were found only in genus 0 up to four point functions \cite{BAlZ}.

The term Matrix model referred to related but different things. For example Gross and Migdal in the classical paper \cite{GM} compute the correlation functions for the $p$-critical matrix model as derivatives at different point $t_0=t_1=\ldots=t_{p-2}=0,\, t_{p-1}=t$. Another possibility is to tend (formally) the number of the parameters to infinity (the potential of the model became not a polynomial but a power series). The free energy function depending in infinitely many variables $F(t_0, t_1, \ldots)$ can be defined as the solution of the string and KdV equations. The Witten's conjecture \cite{W2} states that this function $F(t_0, t_1, \ldots)$ coincides with the Topological gravity generating function. This conjecture was proved by Kontsevich \cite{MK}.

It is natural to ask how to compare these two solutions of the string equation namely the generating function of the Topological gravity $F^{\rm TG}(t_0, t_1, \ldots)$ and the matrix model free energy $F^{\rm MM}(t_0,t_1,\ldots,t_{p-1})$ mentioned in the second paragraph. It is explained in Section 3 that after the naive vanishing of the extra parameters these functions do not coincide. However, these functions are connected by an nontrivial analytic continuation.

The generating function of the Topological gravity $F(t_0, t_1, \ldots)$ satisfies some partial differential equations which are equivalent to the Topological recursion relations (TRR) for intersection numbers on the moduli spaces of Riemann surfaces. These differential equations involve only finitely many variables. Hence, from the analytic continuation property mentioned above follows that TRR hold for the Matrix model function $F^{\rm MM}(t_0,t_1,\ldots,t_{p-1})$. In Appendix B we check these relations in genus 2 by a direct computation. The fulfilment of TRR in this Matrix model was checked for the genus 0 in  \cite{Z}, for the genus 1 in \cite{BT}.

In Appendix A we derive the explicit expressions for low genera free energy in the $p$-critical Matrix model using the Douglas string equation. The concluding formulas are not new and was obtained by Itzykson and Zuber for the "universal one-matrix model" in \cite{IZ} by the different method.

Another recent approach to the relation between the p-critical Matrix models and Topological gravity is given in \cite{EynOran, BerEyn}.

\section{Preliminaries}

\subsection{Liouville gravity}
In this subsection we briefly recall the definition of the correlation functions in the Minimal Liouville gravity. Details can be found in \cite{AlZa, BZ} .

In this paper we need only the $(2,2p+1)$ Minimal Liouville Gravity.  The total action of the Liouville gravity reads
$$
S=S_{\rm L}+S_{\rm Ghost}+S_{\rm MM},
$$
where $S_{\rm MM}$ stands for the $(2,2p+1)$ Minimal CFT action, $S_{\rm Ghost}$ ia a standard ghost action and the Liouville action reads
$$
S_L [\phi] = {1\over{4\pi}}\, \int\,\sqrt{\hat g}\,\,\bigg[{\hat
g}^{\mu\nu}\partial_{\mu}\varphi
\partial_{\nu}\varphi + Q\,{\hat R}\,\varphi + 4\pi\mu\,e^{2b\,\varphi}
\bigg]\,d^2 x,
$$
where $b=\sqrt{2/(2p+1)}$ and the parameter $\mu$ is interpreted as the cosmological constant. The observables are defined as
$$
O_k = \int\,\Phi_{1,{k+1}} (x)\,V_{1,{-k-1}} (x)\,d^2 x,
$$
where $\Phi_{1,\, k+1}$, $V_{1,\, -k-1}(x)$ are certain primary fields of the matter CFT and Liouville theory respectively, $0 \leq k \leq p-1$.
The correlation functions defined by the formula
$$
\langle\, O_{k_1} ...  O_{k_N}\, \rangle = \,\int\, O_{k_1} ...
O_{k_N}\,e^{-S[g,\phi]}\,D[g,\phi ].
$$
It is convenient to define the Liouville gravity generating (or partition) function
\begin{align}
 F^{\rm LG}(\{\lambda\})=\sum\limits_{k_1,k_2,\ldots}\left \langle O_{k_1} ...  O_{k_N} \right\rangle \frac{\lambda_{k_1}\ldots\lambda_{k_n}}{|\mathrm{Aut}(k_1,\ldots,k_n)|}=\int\,D[g,\phi]\,e^{-S_{\lambda} [g,\phi]}, \label{FLG}
\end{align}
$$
S_{\lambda} [g,\phi] = S [g,\phi] + \sum_{k=1}^{p-1}\,\lambda_k\,O_k.
$$
Consider the genus expansion of the function $$F^{\rm LG}=F^{\rm LG}_0+F^{\rm LG}_1+\ldots,$$ where $F^{\rm LG}_g$ is a generating function of the genus $g$ correlation functions. The functions $F^{\rm LG}_g$ have the scaling property e.g. $$F^{\rm LG}_0(\rho^{\delta_0}\mu,\rho^{\delta_k}\lambda_k)=\rho^{\frac{2p+3}2}F^{\rm LG}_0(\mu,\lambda_k).$$
The gravitation scaling dimensions of the variables
\begin{align} \delta_k=\frac{k+2}2. \label{scdmLG} \end{align}
Hence the function $F^{\rm LG}_0$ should has the form
\begin{align} F^{\rm LG}_0=\mu^\frac{2p+3}2 g\left(\frac{t_1}{\mu^{3/2}},\frac{t_2}{\mu^2},\ldots,\frac{t_{p-1}}{\mu^{(p+1)/2}}\right). \label{LiF0}\end{align}
It follows from the definition of the function $F^{\rm LG}$ \eqref{FLG} that the function $g$ is defined by the expansion into the power series.

\subsection{Matrix models}
In this section we give some basic notion on the Matrix models. Details can be found in the reviews \cite{rev1, rev2} or in the recent paper \cite{BT}. The free energy of the one-matrix approach is defined by the matrix integral
$$
F(v_{k},N ) = \log \int dM e^{-\tr V(M)},
$$
where $M$ is a Hermitian $N\times N$ matrix and $V(M)=N\sum v_{k}M^{2k} $ is a polynomial potential. It is known \cite{rev1, rev2} that the function $F$ can be expanded into the series
\begin{align}
F(v_k,N) = \sum_{g=0}^{\infty} N^{2-2g}F_{g}(v_k). \label{FZ}
\end{align}
Each term $F_g$ is equal to the sum of contributions of connected surfaces of genus $g$ made of polygons. The one-matrix model possesses a set of multi-critical points, labeled by integer $p=1,2,3,...$ in the space of the ``potentials'' $V (M) = N\sum v_{k}M^{2k} $. The $p$-critical point exists if the number of variables is greater then $p$ i.e. the degree of $V$ is greater than $2p$. We consider the $p$-critical point for the potential $V(M)=N\sum_{k=1}^{p+1}v_{k}M^{2k}$. The leading singular term of the function $F_g$ has the scaling property
\begin{align} F_g^{sing}[\lambda^{\frac{k+2}{2}}w_{k},N]
=(\lambda^{p+3/2})^{1-g}F^{sing}_{g}[w_{k},N], \label{SCAL} \end{align} where $w_k$ are certain coordinates centred at the $p$-critical point. Double scaling limit corresponds to $N \to \infty$ and $w_k \to 0$ as $w_k=(N^{2}\varepsilon^{2})^{-\frac{k+2}{2p+3}}t_k$. Usually $\ve$ is set to $1$, but we keep $\ve$ as a parameter in order to consider the genus expansion.  In the double scaling limit near the $p$-critical point the expression (\ref{FZ}) looks like
\begin{align}
F^{sing}[w_k] = \sum_{g=0}^{\infty} N^{2-2g}F^{sing}_{g}\left((N^{2}\varepsilon^{2})^{-\frac{k+2}{2p+3}}t_k\right)= \ve^{-2} \sum_{g=0}^{\infty}
\varepsilon^{2g}F^{sing}_{g}[t_{k}],
 \label{Ze}
\end{align}
where we used the scaling property \eqref{SCAL} in the second equality. Below we will denote $F_g^{sing}$ as $F_g$ and perform the substitution $\ve^{-2} F\mapsto F$ for simplicity.

\subsection{The String equation}
The key property of the function $F$ is the fulfilment of the string equation and the KdV equations. Denote by $t_0, t_1,\ldots, t_{p-1}$ the KdV coordinates near the $p$-critical point. All requirements for the KdV coordinates are stated below (for the definition see e.g. \cite[Sec. 4]{BT} or the reviews \cite{rev1, rev2}). Let
$$u(t_0,t_1,\ldots,t_{p-1},\ve)=\pd ^2F(t_0,t_1,\ldots,t_{p-1},\ve) / \pd t_{p-1}^2.$$
Below we use the notation $t_{-2}=1,$ $t_{-1}=0,$ $x=t_{p-1},$ $d=\pd / \pd x.$ The string equation reads
\begin{align}
[\hat{P},\hat{Q}]=\varepsilon, \label{SQ}
\end{align}
where $\hat{Q}=\varepsilon^{2}d^{2}+u$ and $\hat{P}
=-\sum_{k=1}^{p+1}t_{p-1-k}\hat{Q}^{k-1/2}_{+}$ are two differential
operators. $\hat{Q}^{k-1/2}_{+}$ is the non-negative part of the
pseudo-differential operator $\hat{Q}^{k-1/2}$. The function $u(t_0,t_1,\ldots,t_{p-1},\ve)$ is a solution of (\ref{SQ}).

It is known (see e.g. \cite[App. A]{rev2}) that
\begin{align} [\hat{Q}^{k-1/2}_{+},\hat{Q}]=\ve \frac{dR_{k}}{dx}, \label{Rk} \end{align}
where $R_k(u,u_x,u_{xx},\ldots)$ are the Gelfand--Dikii polynomials in $u$ and its $x$ derivatives. These polynomials are determined by the recursion relation
\begin{align}
\frac{dR_{k+1}}{dx}= u\frac{dR_{k}}{dx}+\frac{1}{2}u_{x}R_{k}
+\frac{\varepsilon^{2}}{4}\frac{d^{3}R_{k}}{dx^{3}}, \label{RR}
\end{align}
with the boundary conditions  $R_{1}=u$ and $R_{k}$ vanish at $u=0$. The first polynomials have the form
$$R_1=u, \quad\quad R_2=\frac34u^2+\ve^2\frac14u_{xx},\quad\quad R_3=\frac58u^3+\ve^2\left(\frac58uu_{xx}+\frac5{16}u^2_x\right)+\ve^4\frac1{16}u_{xxxx},$$ $$ R_k=\frac{(2k-1)!!}{2^kk!}u^k+ o(\ve).$$

It follows from (\ref{SQ}) and (\ref{Rk}) that
\begin{align} \sum_{k=1}^{p+1}t_{p-1-k}R_{k}(u) =-  x.  \label{SQ2} \end{align}
We are looking for the solution $u$ in the form
$$ u(t,\ve)=u_0+u_1\ve^2+u_2\ve^4+\ldots $$
By taking the zeroth order of \eqref{SQ2} and rescaling the parameter $t_{p-1-k} \mapsto \frac{2^{k-1}k!}{(2k-1)!!}t_{p-1-k}$ we get
\begin{align} \mathcal{P}(u_0) = \sum_{k=0}^{p+1}t_{p-k-1}u_0^{k} = u_0^{p+1} + \sum_{k=0}^{p-1}t_{p-k-1}u_0^{k} =0 \label{P}.\end{align}
It Appendix A we explain how to find $u_k$ for $k>0$ and give the expressions for $u_1$, $u_2$. It remains to choose the solution of the equation \eqref{P}. These functions in the variables $t_k$ have the scaling property with the dimensions of the variables $$\dim(t_k) = \frac{k+2}2.$$
Comparing with \eqref{scdmLG} we conclude that $t_0$ has the same scale dimension as the cosmological constant $\mu$ in the Liouville theory. These parameters have lowest dimension hence its identification is unique up to scalar multiply. We identify $t_0$ and $-\mu$. Using \eqref{LiF0} we get
\begin{align}
u_0(\mu,t_1,t_2,\ldots,t_{p-1})=\frac{\pd ^2F(t_0,t_1,\ldots,t_{p-1})}{\pd t_{p-1}}=\mu^{1/2}g\left(\frac{t_1}{\mu^{3/2}},\frac{t_2}{\mu^2},\ldots,\frac{t_{p-1}}{\mu^{(p+1)/2}}\right) \label{U*},
\end{align}
where the function $g$ is specified by the expansion into the power series. This choice of the root was mentioned in the Introduction.

\section{Comparison with Topological gravity}

In \cite{W1} Witten introduced the Topological gravity. The correlation functions in this theory are defined in terms of intersection numbers on the moduli spaces of complex curves with marked points.

Let $\mathcal{M}_{g,n}$ be the moduli space of complex curves of the genus $g$ with $n$ ordered marked points and $\overline{\mathcal{M}}_{g,n}$ be its Deligne-Mumford compactification. $\overline{\mathcal{M}}_{g,n}$ is the moduli space of stable curves \footnote{Recall that stable curve means connected, projective curve with no singularities other than double points and with a finite automorphism group. The precise definition and details can be found e.g. in \cite{LZ}}. $\overline{\mathcal{M}}_{g,n}$ is not a manifold but an orbifold (i.e. locally the quotient of a manifold by a finite group). Its complex dimension is $3g-3+n$.

There are natural cohomology classes on such moduli spaces. Let $\Sigma$ be a stable curve with marked points $x_1,\ldots, x_n$. By the definition of
Deligne-Mumford compactification the curve $\Sigma$ may has singularities (double points), but the marked points must be smooth. Thus the cotangent space
$T^{*}_{x_k}\Sigma$ is well defined and the holomorphic bundles $\mathcal{L}_k$ on $\overline{\mathcal{M}}_{g,n}$ with the fiber $T^{*}_{x_k}\Sigma$ over
$(\Sigma,x_1,\ldots,x_n)$ can be defined as well. Denote by $\psi_k$ the first Chern class of $\mathcal{L}_k$, $\psi_k=c_1(\mathcal{L}_k).$

The correlation numbers in the Topological gravity have the form
\begin{align}
\left\langle \tau_{d_1}  \cdots \tau_{d_n} \right\rangle= \int_{\overline{\mathcal{M}}_{g,n}}\psi_1^{d_1}\cdots\psi_n^{d_n}, \label{in}
\end{align}
where the genus $g$ is uniquely determined by the condition $$d_1+\cdots+d_n=\dim\overline{\mathcal{M}}_{g,n}=3g-3+n.$$
The generating function of these correlation numbers reads\footnote{We use the notation $\Fi,\ti_k,\ui,\Ri_k$ instead of standard $F,t_k,u,R_k$ in order to distinguish the related but not coincident the Topological gravity and the Matrix model objects.}
\begin{align}\Fi(\ti_0,\ti_1,\ldots)= \sum\limits_{d_1,d_2,\ldots}\left \langle \tau_{d_1}  \cdots \tau_{d_n} \right\rangle \frac{\ti_{d_1}\ldots\ti_{d_n}}{|\mathrm{Aut}(d_1,\ldots,d_n)|}.\label{FI}\end{align}

Witten conjectured in \cite{W1} the equivalence between two approaches to 2D Quantum gravity namely the Topological gravity and the Matrix models. The precise formulation
of this conjecture (see e.g. \cite{W2}) states that $\Fi(\ti_0,\ti_1,\ldots)$ satisfies the KdV hierarchy
$$\frac{\pd \ui}{\pd \ti_{k}} = \frac{\pd \Ri_{k+1}[\ui]}{\pd \ti_0}$$
and the string equation
$$\frac{\pd \Fi}{\pd \ti_0}= \frac{\ti_0^2}2+\sum\limits_{k=0}^{\infty} \ti_{k+1} \frac{\pd \Fi}{\pd \ti_{k}}.$$
Here $\ui=\pd^2 \Fi/\pd \ti_0^2$ and each polynomials $\Ri_k$ will coincide with $R_k/2(2k-1)!!$ if we set $u=2\ui$ and $\ve=1$. These polynomials $\Ri_k$ have the form
$$\Ri_1=\ui, \quad\quad \Ri_2=\frac{\ui^2}2+\frac{\ui_{xx}}{12}, \quad\quad \Ri_3=\frac{\ui^3}6+\frac{\ui\ui_{xx}}{12}+\frac{\ui^2_x}{24}+\frac{\ui_{xxxx}}{240}, \quad\ldots\quad \Ri_k=\frac{\ui^k}{k!}+\ldots$$
This conjecture was proved by Kontsevich \cite{MK}. The string and the KdV equations appear in the Matrix models as an equations for the free energy function.

It is natural to ask how to relate the generating function $\Fi(\ti_0,\ldots)$ in infinitely many variables to the free energy $F(t_0,\ldots,t_{p-1})$ of the $p$-critical Matrix model. For example, assume that we know all intersection numbers (\ref{in}), how to find the power series expansion of the function $g$ in (\ref{U*})?

Consider the simplest example $p=2$. Using the KdV equations, we get from the string equation
$$\frac{\pd \ui}{\pd \ti_0}= 1+\sum\limits_{k=0}^{\infty} \ti_{k+1} \frac{\pd \Ri_{k+1}[\ui]}{\pd \ti_0}.$$
Integrating and restricting it to the genus 0 part, we obtain
$$\sum\limits_{k=0}^{\infty} \ti_{k}\frac{\ui_0^k}{k!}-\ui_0=0.$$
Comparing this with the genus 0 string equation in the Matrix models (\ref{P0}) in the $p=2$ case we relate the variables
\begin{align} \ti_0=t_1=x, \quad \ti_1-1=t_0=-\mu, \quad \ti_3=3!, \quad \ti_2=\ti_4=\ti_5=\cdots=0.\label{TIT}\end{align}
Substituting (\cite[Prop 4.6.10]{LZ})
$$\left\langle \tau_{d_1}  \cdots \tau_{d_n} \right\rangle_0  = \frac{(n-3)!}{d_1!\cdots d_n!}.$$
in \eqref{FI}, we obtain the genus 0 generating function
$$\Fi_0(\ti_0,\ti_1,\ldots)= \sum\limits_{\substack{k_1,k_2,\ldots \\ k_0=k_2+2k_3+3k_4+\ldots+3}} \frac{(k_1+2k_2+3k_3+\ldots)!}{0!^{k_0}1!^{k_1}\cdots} \cdot\frac{\ti_{0}^{k_0}\ti_{1}^{k_1}\cdots}{k_0!k_1!\cdots}.$$
With the relation (\ref{TIT}), we get
$$\ui_0(\ti_0,\ti_1)= \sum\limits_{\substack{k_1,k_3 \\ k_0=2k_3+3}} \frac{(k_1+3k_3)!}{0!^{k_0}1!^{k_1}(3!)^{k_3}} \cdot\frac{\ti_{0}^{2k_3+1}\ti_{1}^{k_1} (3!)^{k_3}}{(2k_3+1)!k_1!k_3!}=\sum\limits_{k_1,k_3} \frac{(k_1+3k_3)!}{(2k_3+1)!k_1!k_3!} \ti_{0}^{2k_3+1}\ti_{1}^{k_1}.$$
Finally, summing on $k_1$, we obtain
\begin{align}\ui_0=\sum\limits_{k_3} \frac{(3k_3)!}{(2k_3+1)!k_3!} \frac{\ti_{0}^{2k_3+1}}{(1-\ti_1)^{3k_3+1}}=\frac{x}{\mu}+\frac{x^3}{\mu^4}+3\frac{x^5}{\mu^7}+\cdots \label{Ui0p2} \end{align}
This function differs from the Matrix models $u$ in formula (\ref{U*})
\begin{align} u_0=\mu^{1/2}g\left(\frac{x}{\mu^{3/2}}\right)=\mu^{1/2}-\frac{x}{2\mu}-\frac{3x^2}{8\mu^{5/2}}+\cdots \label{U0p2} \end{align}

Let us consider the expansions \eqref{Ui0p2} and \eqref{U0p2} of the roots of the string equation $u^3-\mu u +x=0$ at the point $x=0.$ At this point \eqref{Ui0p2} and \eqref{U0p2} equal to $0$ and $\mu^{1/2}$ respectively. A root of an algebraic equation is locally given by an analytical function in the coefficients of the equation. If the coefficients run round the discriminant set\footnote{The set where the discriminant of the corresponding polynomial vanishes} the roots of equation permute. The group of such permutations is known as the monodromy group. In this case this group is a Galois group of general polynomial and equals to the group of all permutations. Thus \eqref{Ui0p2} and \eqref{U0p2} do not coincide but are connected by a nontrivial analytic continuation in $x$.

It was explained in Appendix A that $u_g$ for $g>0$ can be expressed as rational functions in $u^*=u_0$ and its $x$ derivatives. The argument there explores only
the string equation, hence, \eqref{U1} and \eqref{U2} should be satisfied in the Topological gravity. Therefore, not only $\ui_0$ transfers to $u_0$ by an analytic continuation but the whole $\ui=\sum \ve^{2k}\ui_k$
transfers to $u= \sum \ve^{2k}u_k$.

For $p>2$ the argument is quite similar. In the general case $u$ and $\ui$ are expansions of roots of the string equation $\mathcal{P}(u_0)=0$ at different points unlike the $p=2$ case. Still they are connected by an analytic continuation in the variables $t_0,t_1,\ldots$

\section{Acknowledgements}

We are grateful to O.~Bershtein, B.~Feigin, S.~Lando, M.~Lashkevich, A.~Marshakov, A.~Mironov  and A.~Morozov  for useful discussions.
We are also grateful to L.~Chekhov, B.~Eynard and I.~Kostov for careful reading the first version of the article and useful remarks.

This research was carried out in the framework of the Federal
Program ``Scientific and Scientific-Pedagogical personnel of
innovational Russia'' No 1339, RFBR initiative interdisciplinary
project 09-02-12446-ofi-m and RFBR-CNRS project PICS-09-02-91064.

\section*{Appendix}
\appendix
\section{}

In this Appendix we evaluate the low genera free energy using the Douglas string equation \eqref{SQ2}. The concluding formulas \eqref{F1}, \eqref{F2} are not new and was obtained by Itzykson and Zuber by another method for the "universal one-matrix model" in \cite[Sec. 5]{IZ}.

We are looking for $u$ in the form
\begin{align} u(t,\ve)=u_0+u_1\ve^2+u_2\ve^4+\ldots \label{uex} \end{align}
It is natural to study the $\ve$ expansion of $R_k$ first. Two properties follow from the recursion relation (\ref{RR}) by induction argument

1. Only even degrees of $\ve$ appear in $R_k$. Hence, $R_k$ can be expanded as
$$R_k=R_k^0+R_k^1\ve^2+R_k^1\ve^4+\ldots$$
For each $k$ this sum is finite.

2. The polynomial $R_k^{l}$ is a linear combination of monomials, which have $2l$ derivatives with respect to~$x$ each. The degree of $R_k^{l+1}$ is one less than the degree of $R_k^{l}$. Moreover $\deg R_k^l=k-l$. Hence,
$$R_k^0=A_ku^k,$$
$$R_k^1=B_k^1u^{k-2}u_{xx}+B_k^2u^{k-3}u_x^2,$$
$$R_k^2=C_k^{1}u^{k-3}u_{xxxx}+C_k^{2}u^{k-4}u_{xxx}u_{x}+C_k^{3}u^{k-4}u_{xx}^{2}
+C_k^{4}u^{k-5}u_{x}^{2}u_{xx}+C_k^{5}u^{k-6}u_{x}^{4}.$$
Now (\ref{RR}) leads to a system of linear recursion relations on $A, \{B\},\{C\},\ldots$ This system has upper triangular form and can be easily solved.
The answer reads

\begin{align}
R_{k}(u) =&\frac{(2k-1)!!}{2^{k-1}k!}\cdot
\left(u^{k}+\ve^{2}\left[\frac16(u^{k})''u_{xx}+\frac1{12}(u^{k})'''u_{x}^{2}\right]
+\ve^{4}\left[\frac{1}{60}(u^{k})'''u_{xxxx}+\frac{1}{30}(u^{k})^{(4)}u_{xxx}u_{x}+\right.\right.\notag\\
&\left.\left.+\frac{1}{40}(u^{k})^{(4)}u_{xx}^{2}
+\frac{11}{360}(u^{k})^{(5)}u_{x}^{2}u_{xx}+\frac{1}{288}(u^{k})^{(6)}u_{x}^{4}\right]+O(\ve^{6}) \right), \label{SQF}
\end{align}
where $'$ stands for the $u$ derivative (e.g. $(u^{k})''=k(k-1)u^{k-2}$). Note that $k$ dependance is supported only in the common factor and
the $u$ derivatives. This feature can be obtained from the identity (\cite[App. A]{rev2})
\begin{align}\frac{\pd}{\pd u}R_{k}(u)=\frac{2k-1}2R_{k-1}(u). \label{RR2} \end{align}
Indeed, consider polynomials of the form $$H_k=\frac{(2k-1)!!}{2^k\cdot k!}(u^k)^{(l)}h(u_x,u_{xx},\ldots),$$ where $h(u_x,u_{xx},\ldots)$ does not depend on $k$.
They are solutions of the equation (\ref{RR2}). Hence, $R_k$ is a linear combination of such polynomials with constant coefficients. We use this feature below (see \eqref{DSF}).

Now we can find $u_0, u_1, u_2.$
By substituting (\ref{SQF}) into this equation and rescaling the parameters $t_{p-1-k} \mapsto
\frac{2^{k-1}k!}{(2k-1)!!}t_{p-1-k}$ we get
\begin{align}
&\mathcal{P}(u)+\varepsilon^{2}\left[\frac{1}{6}\mathcal{P}''(u)u_{xx}
+\frac{1}{12}\mathcal{P}'''(u)u_{x}^{2}\right]+
\varepsilon^{4}\left[\frac{1}{60}\mathcal{P}^{(3)}(u)u_{xxxx}+\frac{1}{30}\mathcal{P}^{(4)}(u)u_{xxx}u_{x}+\right.  \notag\\
&+\left.\frac{1}{40}\mathcal{P}^{(4)}(u)u_{xx}^{2}+\frac{11}{360}\mathcal{P}^{(5)}(u)u_{x}^{2}u_{xx}
+\frac{1}{288}\mathcal{P}^{(6)}(u)u_{x}^{4}\right]=O(\ve^{6}), \label{DSF}
\end{align}
where
$$
\mathcal{P}(u) = \sum_{k=0}^{p+1}t_{p-k-1}u^{k} = u^{p+1}+\sum_{k=0}^{p-1}t_{p-k-1}u^{k}
.$$

By taking the zeroth order of \eqref{DSF} with $u$ of the form \eqref{uex}, we get
\begin{align} \cP(u_{0})=0. \label{P0}\end{align}
Therefore
\begin{align}
u_{0}=u^{*}(t_{0},...,t_{p-2},x), \label{U0}
\end{align}
where $u^{*}$ is a suitably chosen root of the polynomial
$\cP(u)$ (See the Subsection 2.3).  The second and fourth orders in the $\varepsilon$ expansion give
\begin{align}
u_{1}=&-
\frac{2u_{xx}^{*}\mathcal{P}_2+(u^{*}_{x})^{2}\mathcal{P}_3}{12\mathcal{P}_1},
\label{U1} \\
u_{2}=&-\frac{\left[\frac{1}{60}u_{xxxx}^{*}\mathcal{P}_3+\frac{1}{30}u_{xxx}^{*}u_{x}^{*}\mathcal{P}_4
+\frac{1}{40}(u_{xx}^{*})^{2}\mathcal{P}_4+\frac{11}{360}(u_{x}^{*})^{2}u_{xx}^{*}\mathcal{P}_5
+\frac{1}{288}(u_{x}^{*})^{4}\mathcal{P}_6\right]}{\mathcal{P}_1}-\notag\\
&-\frac{\left[\frac{1}{6}(u_{1})_{xx}\mathcal{P}_2+\frac{1}{6}u_{1}u_{xx}^{*}\mathcal{P}_3
+\frac{1}{6}(u_{1})_{x}u_{x}^{*}\mathcal{P}_3+\frac{1}{12}
u_{1}(u_{x}^{*})^{2}\mathcal{P}_4\right]}{\mathcal{P}_1}-\frac{u_{1}^{2}\mathcal{P}_2}{2\mathcal{P}_1}, \label{U2}
\end{align}
where $\mathcal{P}_k$ stands for $\mathcal{P}^{(k)}(u^*)$.

This procedure can be continued. It is easy to see that for any $g$ the $u_g$ can be expressed as a rational function in $u^{*}$, it's $x$ derivatives and $\mathcal{P}^{(k)}(u^*)$.

It remains to find $F_0, F_1, F_2$ from $u_0,u_1,u_2$. Since
$$ \frac{\partial^{2} F_g}{\partial x^{2}} = u_g= f_g(u^{*}), $$
we have
$$ F_g = -\int_{0}^{u^{*}}\cP(u)\cP'(u)f_g(u)du. \label{ans1}$$
This formula can be checked by a straightforward calculation. Integrating by parts and omitting the regular terms, we get from
(\ref{U0}), (\ref{U1}) and (\ref{U2})
\begin{align}
F_{0}&=
 \frac{1}{2}\int_{0}^{u^{*}}\cP^{2}(u)du, \label{F0}\\
F_{1}&=-\frac{\log \cP'(u^{*})}{12}, \label{F1} \\
F_{2}&=-\frac{1}{1440}\left(5\frac{\mathcal{P}^{(4)}(u^{*})}{(\mathcal{P}')^{3}}
-29
\frac{\mathcal{P}''(u^{*})\mathcal{P}'''(u^{*})}{(\mathcal{P}')^{4}}
+28\frac{(\mathcal{P}''(u^{*}))^{3}}{(\mathcal{P}')^{5}}\right). \label{F2}
\end{align}

The genus $g$ correlation functions can be evaluated by the formula
\begin{align}
\langle O_{k_{1}}...O_{k_{n}}\rangle_{g}  = \frac{\partial^{n}
F_{g}}{\partial t_{p-k_{1}-1}\ldots\partial t_{p-k_{n}-1}}.
\label{cf} \end{align}
The operators $O_k$ correspond to the operators $O_{p-1-k}$ in the notation of \cite{BZ, BT}. Using \eqref{F0}, \eqref{F1}, \eqref{F2} one can get the correlation functions at genera $0, 1, 2$:
\begin{align}
&\langle O_{k_{1}}O_{k_{2}}\rangle_{0} =
\frac{(u^{*})^{k_{1}+k_{2}+1}}{k_{1}+k_{2}+1},\qquad  \langle
O_{k_{1}}O_{k_{2}}O_{k_{3}}\rangle_{0} =
-\frac{(u^{*})^{k_{1}+k_{2}+k_{3}}}{\mathcal{P}_{1}}, \notag\\
&\langle O_{k_{1}}O_{k_{2}}O_{k_{3}}O_{k_{4}}\rangle_{0}=\frac{k(u^{*})^{k-1}}{\mathcal{P}_{1}^{2}}
-\frac{\mathcal{P}_{2}}{\mathcal{P}_{1}^{3}}(u^{*})^{k},\notag\\
&\langle O_{k_{1}}O_{k_{2}}O_{k_{3}}O_{k_{4}}O_{k_{5}}\rangle_{0}
=-\frac{k(k-1)}{\mathcal{P}_{1}^{3}}(u^{*})^{k-2}+\frac{3k\mathcal{P}_{2}}{\mathcal{P}_{1}^{4}}(u^{*})^{k-1}
+\left(\frac{\mathcal{P}_{3}}{\mathcal{P}_{1}^{4}}-\frac{3\mathcal{P}_{2}^{2}}{\mathcal{P}_{1}^{5}}\right)(u^{*})^{k}.
\label{g0cn}
\end{align}

\begin{align}
&\langle O_{k} \rangle_{1} =
-\frac{k(u^{*})^{k-1}}{12\mathcal{P}_{1}}+\frac{\mathcal{P}_{2}(u^{*})^{k}}{12\mathcal{P}_{1}^{2}}
, \notag \\
&\langle O_{k_{1}} O_{k_{2}} \rangle_{1} =
\frac{k_{1}^{2}+k_{2}^{2}-k_{1}-k_{2}+k_{1}k_{2}}{12\mathcal{P}_{1}^{2}}(u^{*})^{k-2}
-\frac{k\mathcal{P}_{2}}{6\mathcal{P}_{1}^{3}}(u^{*})^{k-1}+\left(\frac{\mathcal{P}_{2}^{2}}{6\mathcal{P}_{1}^{4}}
-\frac{\mathcal{P}_{3}}{12\mathcal{P}_{1}^{3}}\right)(u^{*})^{k},
\label{g1cn}
\end{align}

\begin{align}
\langle O_{k}\rangle_{2} =& -\frac{1}{1440}\left[
\frac{5k(k-1)(k-2)(k-3)}{\mathcal{P}_{1}^{3}}(u^{*})^{k-4}-
\frac{29k(k-1)(k-2)\mathcal{P}_{2}}{\mathcal{P}_{1}^{4}}(u^{*})^{k-3}+\right.
\notag \\
&+\left.
k(k-1)\left(\frac{84\mathcal{P}_{2}^{2}}{\mathcal{P}_{1}^{5}}
-\frac{29\mathcal{P}_{3}}{\mathcal{P}_{1}^{4}}\right)(u^{*})^{k-2}+k\left(\frac{116\mathcal{P}_{2}\mathcal{P}_{3}}{\mathcal{P}_{1}^{5}}-
\frac{140\mathcal{P}_{2}^{3}}{\mathcal{P}_{1}^{6}}
-\frac{15\mathcal{P}_{4}}{\mathcal{P}_{1}^{4}}\right)(u^{*})^{k-1}+
\right.\notag\\
&\left.
+\left(-\frac{5\mathcal{P}_{5}}{\mathcal{P}_{1}^{4}}+\frac{44\mathcal{P}_{2}\mathcal{P}_{4}}
{\mathcal{P}_{1}^{5}}+\frac{29\mathcal{P}_{3}^{2}}{\mathcal{P}_{1}^{5}}
-\frac{200\mathcal{P}_{2}^{2}\mathcal{P}_{3}}{\mathcal{P}_{1}^{6}}
+\frac{140\mathcal{P}_{2}^{4}}{\mathcal{P}_{1}^{7}}\right)(u^{*})^{k}\right],
\label{g2cn}
\end{align}
where $\mathcal{P}_k$ stands for $\mathcal{P}^{(k)}(u^*)$.

\section{}
The intersection numbers (\ref{in}) in genus 0,1 satisfy Topological recursion relations (TRR). These relations reflect the fact that for $g=0,1$ the homology class dual to $\psi_1$ is the boundary class on $\overline{\mathcal{M}}_{g,n}$ i.e. its restriction to $\mathcal{M}_{g,n}$ vanishes. Let
$$\la\la\tau_{k_1}  \cdots \tau_{k_n} \ra\ra=\frac{\pd^n \Fi }{\pd \ti_{k_1}\ldots \pd \ti_{k_n}}=\sum\limits_{d_1,d_2,\ldots}\left \langle \tau_{k_1} \cdots \tau_{k_n}\tau_{d_1}  \cdots \tau_{d_k} \right\rangle \frac{\ti_{d_1}\ldots\ti_{d_n}}{|\mathrm{Aut}(d_1,\ldots,d_n)|}$$
denote the correlation functions of the Topological gravity. The correlation functions at given genus $\la\la\cdot\ra\ra_g$ are defined similarly as a derivatives of $\Fi_g$. The TRR have the form
\begin{align}&\textrm{genus 0:} \qquad \la\la
\tau_{k_{1}}\tau_{k_{2}} \tau_{k_{3}} \ra\ra_{0} =
\la\la \tau_{k_{1}-1}\tau_{0} \ra\ra_{0}\la\la
\tau_{0}\tau_{k_{2}}\tau_{k_{3}} \ra\ra_{0}, \notag\\
&\textrm{genus 1:} \qquad \la\la \tau_{k} \ra\ra_{1}
= \frac{1}{24} \la\la \tau_{k-1} \tau_{0}\tau_{0}
\ra\ra_{0} + \la\la \tau_{k-1} \tau_{0}
\ra\ra_{0}\la\la \tau_{0} \ra\ra_{1}. \notag\end{align}
For the genus 2 Getzler \cite{Ge} proved two recursion relations based on the fact that the homology classes dual to $\psi_1^2$ and $\psi_1\psi_2$ are the boundary classes on $\overline{\mathcal{M}}_{2,n}$. The first relation has the form
\begin{align}\textrm{genus 2:} \qquad &\la\la  \tau_{k}\ra\ra_{2}
= \la\la \tau_{k-1}\tau_{0}\ra\ra_{0}\la\la
\tau_{0}\ra\ra_{2}+\la\la
\tau_{k-2}\tau_{0}\ra\ra_{0}\bigg(\la\la
\tau_{1}\ra\ra_{2}-\la\la \tau_{0}\tau_{0}\ra\ra_{0}\la\la
\tau_{0}\ra\ra_{2}\bigg)+\notag \\
&+\la\la  \tau_{k-2}
\tau_{0} \tau_{0}\ra\ra_{0}\left(\frac{7}{10}\la\la
\tau_{0}\ra\ra_{1}\la\la
\tau_{0}\ra\ra_{1}+\frac{1}{10}\la\la
\tau_{0}\tau_{0}\ra\ra_{1}\right)+
\frac{13}{240}\la\la
\tau_{k-2}\tau_{0}\tau_{0}\tau_{0}\ra\ra_{0}\la\la
\tau_{0}\ra\ra_{1}-\notag\\ &-\frac{1}{240}\la\la
\tau_{k-2}\tau_{0}\ra\ra_{1}\la\la
\tau_{0}\tau_{0}\tau_{0}\ra\ra_{0}+ \frac{1}{960}\la\la
\tau_{k-2}\tau_{0}\tau_{0}\tau_{0}\tau_{0}\ra\ra_{0}. \notag
\end{align}
The second Getzler relation is twice longer and is given in \cite[eq. (7)]{Ge}.
Each relation can be considered as a partial differential equation for the function $\Fi$ e.g. the genus 0 relation can be reduced to the form
$$\frac{\pd^3 \Fi_0}{\pd \ti_{k_1} \pd \ti_{k_2} \pd \ti_{k_2}}=\frac{\pd^2 \Fi_0}{\pd \ti_{k_1-1} \pd \ti_{k_0} }\frac{\pd^3 \Fi_0}{\pd \ti_{0} \pd \ti_{k_2} \pd \ti_{k_2}}.$$
The free energy $F$ of the $p$-critical Matrix model is an analytic continuation of the generating function $\Fi$ of the Topological gravity after vanishing some of the variables $\ti_k$ (e.g. (\ref{TIT})).
Since the TRR involve only finitely many $\ti_k$, the function $F$ should satisfy TRR. This argument is indirect but using (\ref{g0cn}), (\ref{g1cn}), (\ref{g2cn}), these relations can be straightforwardly verified. After necessary replacements $\tau_k \leftrightarrow O_{k}/k!$  and $\la\la\; \ra\ra_{g}^{TG} \to \frac{1}{2^{g}}\langle\;
 \rangle_{g}^{MM}$ TRR read:

\begin{align}
&\textrm{genus 0:} \qquad \langle O_{k_{1}}O_{k_{2}}
O_{k_{3}} \rangle_{0} = k_{1}\langle O_{k_{1}-1}O_{0}
\rangle_{0}\langle
O_{0}O_{k_{2}}O_{k_{3}} \rangle_{0},  \notag \\
&\textrm{genus 1:} \qquad \langle O_{k} \rangle_{1} =
\frac{1}{12}k \langle O_{k-1} O_{0}O_{0} \rangle_{0} + k\langle
O_{k-1} O_{0}
\rangle_{0}\langle O_{0} \rangle_{1},\notag \\
&\textrm{genus 2:} \qquad \langle  O_{k}\rangle_{2} =
k\langle O_{k-1}O_{0}\rangle_{0}\langle
O_{0}\rangle_{2}+k(k-1)\langle O_{k-2}O_{0}\rangle_{0}\bigg(\langle
O_{1}\rangle_{2}-\langle O_{0}O_{0}\rangle_{0}\langle
O_{0}\rangle_{2}\bigg)+\notag \\
&\qquad\qquad\qquad\qquad\qquad\quad+k(k-1)\langle  O_{k-2} O_{0}
O_{0}\rangle_{0}\left(\frac{7}{10}\langle O_{0}\rangle_{1}\langle
O_{0}\rangle_{1}+\frac{1}{5}\langle
O_{0}O_{0}\rangle_{1}\right)+\notag\\
&\qquad\qquad\qquad\qquad\qquad\quad+\frac{13}{120}k(k-1)\langle
O_{k-2}O_{0}O_{0}O_{0}\rangle_{0}\langle
O_{0}\rangle_{1}-\notag\\
&\qquad\qquad\qquad\qquad\qquad\quad-\frac{1}{120}k(k-1)\langle
O_{k-2}O_{0}\rangle_{1}\langle
O_{0}O_{0}O_{0}\rangle_{0}+\notag\\
&\qquad\qquad\qquad\qquad\qquad\quad+ \frac{1}{240}k(k-1)\langle
O_{k-2}O_{0}O_{0}O_{0}O_{0}\rangle_{0}. \label{RRMM}
\end{align}
Is easy to see that all these recursion relations are
fulfilled. We also have checked the second Getzler relation.

\end{document}